\documentclass[runningheads]{llncs}
\usepackage[T1]{fontenc}
\usepackage{graphicx}
\usepackage{amsmath, amssymb,bm}
\usepackage{tikz}
\usepackage{pgfplots}
\usepackage{mathtools}
\usepackage{ifthen}

\def\R{{\mathbb{R}}}

\usepackage{placeins}

\newcommand\Tstrut{\rule{0pt}{2.6ex}}         % = `top' strut
\begin{document}
\title{A Wasserstein GAN for Joint Learning\\ of Inpainting and Spatial 
	Optimisation}
\titlerunning{WGAN for Joint Spatial Optimisation and Inpainting}
\author{Pascal Peter}
\authorrunning{P. Peter}
\institute{Mathematical Image Analysis Group,
    Faculty of Mathematics and Computer Science,\\ Campus E1.7,
    Saarland University, 66041 Saarbr\"ucken, Germany.\\	
    \email{peter@mia.uni-saarland.de}}
\maketitle              % typeset the header of the contribution
\begin{abstract}
%The abstract should briefly summarize the contents of the paper in
Image inpainting is a restoration method that reconstructs missing 
image parts. However, a carefully selected mask of known pixels that yield 
a high quality inpainting can also act as a sparse image representation. This 
challenging spatial optimisation problem is essential for practical 
applications such as compression. So far, it has been almost exclusively 
addressed by model-based approaches. First attempts with neural networks seem 
promising, but are tailored towards specific inpainting operators or require 
postprocessing.
To address this issue, we propose the first generative adversarial network 
(GAN) for spatial inpainting data optimisation. In contrast to previous 
approaches, it allows joint training of an inpainting generator and a 
corresponding mask optimisation network. With a Wasserstein distance, we ensure 
that our inpainting results accurately reflect the statistics of natural 
images. This yields significant improvements in visual quality and speed over 
conventional stochastic models. It also outperforms current spatial 
optimisation networks.

\end{abstract}
\section{Introduction}
\label{sec:introduction}

Image inpainting was originally introduced to restore missing or damaged image 
parts \cite{BCMS14,MM98a}. In this classical setting, the known image data is 
predetermined. However, given the original image, one can instead consider a 
spatial optimisation problem: finding a fraction of known data that allows a 
high quality reconstruction of the image with inpainting. These sparse 
representations, the so-called inpainting masks, have practical applications 
such as inpainting-based compression \cite{GWWB08,SPME14} and adaptive 
sampling \cite{DCPC19}. Therefore, many sophisticated model-based approaches 
have been proposed for spatial optimisation of inpainting data
\cite{BBBW08,BLPP16,CRP14,CW21,HSW13,KBPW18,MHWT12}. However, 
due to the unique challenges of this problem, solutions are often 
slow, complicated, or limited to specific inpainting operators.

Recently, some first attempts were made to solve the mask optimisation problem 
with  neural networks \cite{APW21,DCPC19}. However, these approaches have 
limitations. From classical methods it is well known that optimal positions 
for an inpainting mask heavily depend on the inpainting operator \cite{PHNH16}. 
Despite this close connection, existing deep learning approaches do not allow 
to train a pair of inpainting and spatial optimisation networks, but either 
train them separately \cite{DCPC19} or do not allow learned inpainting at all 
\cite{APW21}. 

%............................................................................

\subsection{Our Contribution}
We propose the first generative adversarial approach for deep inpainting and 
spatial optimisation. It consists of three networks: an inpainting generator, a 
mask generator, and a discriminator. The discriminator allows our learned 
inpainting to approximate the statistics of natural images in terms of a
Wasserstein distance, leading to convincing visual quality. Our mask network is 
the first to generate binary inpainting masks directly. It solves 
non-differentiability issues with approaches from neural network-based image 
compression. The combination of these ingredients makes effective joint 
learning of inpainting and mask optimisation possible.

%............................................................................

\subsection{Related Work} 
\label{sec:related}

The selection of suitable known data is highly dependent on the inpainting 
operator. Only for individual operators such as homogeneous diffusion 
\cite{Ii62}, true optimality statements have been proven \cite{BBBW08}, but 
even those can only be approximated in practice. Optimal control 
\cite{BLPP16,CRP14,HSW13} approaches and a recent finite element method 
\cite{CW21} offer good results, but are limited to certain operators. In our 
comparisons, we consider \emph{probabilistic sparsification} (PS) and 
\emph{non-local pixel exchange} (NLPE) \cite{MHWT12} as representatives for 
classical methods. PS is a stochastic greedy  method that gradually removes 
pixels which increase the inpainting error the least. NLPE is a postprocessing 
step which moves mask points to the most promising positions of a randomly 
chosen candidate set. 
Together, they belong to the current state of  the art in quality and are 
applicable to any deterministic inpainting operator. For a more detailed review 
of model-based spatial optimisation, we refer to Alt et al.~\cite{APW21}.

To the best of our knowledge, only two deep learning approaches for  spatial 
inpainting data optimisation exist so far. The network of Alt et al. 
\cite{APW21} differs fundamentally from our approach in that it optimises masks 
for homogeneous diffusion inpainting, not for a deep inpainting. During 
training, the mask network feeds a non-binary confidence map for known data to 
a surrogate network approximating homogeneous diffusion.
It requires postprocessing by stochastic sampling  to obtain the final binary
masks. The adaptive sampling contribution of Dai et al.~\cite{DCPC19} is closer 
in spirit to our approach: It combines a mask network \emph{NetM} with a 
pre-trained inpainting network \emph{NetE}. The authors note that joint 
training of NetM and NetE did not yield satisfying results due the non-binary 
output of NetM. We address this in more detail in Section \ref{sec:binary}.

A full review of the numerous deep inpainting approaches is beyond the scope of 
this paper. Most of these  
\cite{LRSW+18,LJXY19,PKDD16,WZZ21,WZNL+21,XXC12,YLYS+19,WGJW+20} focus on 
classical inpainting problems: Regular shaped regions like squares, circles, 
text, or free form scribbles are removed from the image. Typically this means 
that only a modest amount of data is missing ($10\%$--$60\%$). In contrast, 
sparse spatial optimisation is mostly concerned with much higher amounts of 
unknown data ($>90\%$) since those are interesting for compression or adaptive 
sampling purposes. Moreover, optimised known data is often not only extremely 
sparse, but also does not provide nicely connected regions. Most existing 
approaches are thus not directly applicable and at the very least, the training 
procedure must be adapted. 

Deep learning methods specifically designed for sparse data are much more rare 
\cite{UVL18,VHF21}.
We explain in more detail in Section~\ref{sec:WassersteinGANs}
why we specifically choose Wasserstein GANs \cite{ACB17,VHF21} as a foundation 
for our approach. Note that none of the aforementioned pure inpainting methods 
provides the option for data optimisation or has been previously evaluated on 
optimised known data.

\subsection{Organisation of the Paper} After a brief review of Wasserstein GANs 
in Section~\ref{sec:WassersteinGANs} we introduce our deep spatial optimisation 
approach in Section~\ref{sec:method} and evaluate it in 
Section~\ref{sec:experiments}. The paper concludes with  a discussion and 
outlook on future work in Section~\ref{sec:conclusion}.

%............................................................................

\section{Inpainting with Wasserstein GANs}
\label{sec:WassersteinGANs}

For our data optimisation, we require deep inpainting that is suitable 
for sparse known data. Va\v{s}ata et al.~\cite{VHF21} have successfully 
applied \emph{Wasserstein generative adversarial networks} (WGANs) for
inpainting on random sparse data. Since WGANs are also mathematically 
well-founded, they are a natural starting  point for our approach. In 
particular, \emph{generative adversarial networks} (GANs) \cite{GPMX+14} can be 
seen as generalisation of classical inpainting techniques that achieve high 
quality~\cite{SPME14} by accurately approximating the statistics of natural 
images \cite{PW15a}. 

A GAN relies on two competing networks to generate samples from a 
target distribution $\mathbb{P}_t$. The \emph{generator} takes a sample from a 
source distribution $\mathbb{P}_s$ and maps it to a representative of 
$\mathbb{P}_t$. In our case, $\mathbb{P}_s$ is a uniformly random distribution, 
and $\mathbb{P}_t$ corresponds to the statistics of natural images.
The \emph{discriminator} judges how well the generated representative 
fits to the target distribution. This creates a minmax problem, in which the 
generator tries to trick the discriminator in accepting its result as a true 
sample of $\mathbb{P}_t$.

Unfortunately, GANs tend to suffer from training instabilities due to 
imbalances between the generator and the discriminator. Arjovsky et 
al.~\cite{ACB17} have shown the large impact of the loss function, which 
measures the difference between generator samples and target distributions. 
Using a Wasserstein distance \cite{Va69} instead of the classical 
Jensen-Shannon divergence \cite{GPMX+14} stabilises training, avoids vanishing 
gradients, and indicates training progress more reliably.

Assume we want to inpaint an image of resolution $m \cdot n$ with $k$ 
channels with a WGAN~\cite{VHF21}. We write its $N\!\!:=\!\!mnk$ pixel values 
in vector notation as $\bm 
f \in \R^N$. Data is known at locations where the confidence function $\bm c 
\in [0,1]^N$ is non-zero, thus providing side information for the generator 
$\bm g: {\left(\R^N\right)}^3 \rightarrow \R^N $, a parametric function 
represented by a network.
Representing the known data as $\bm C \bm f$ with a masking matrix $\bm 
C:=\textnormal{diag}(\bm c) \in \R^{N \times N}$, the generator creates the 
inpainting result $\bm u \in \R^N$ based on the inpainting constraint
\begin{equation}
	\bm u(\bm r, \bm c, \bm C \bm f) := (\bm I - \bm C) \bm g(\bm r, \bm c, \bm 
	C 
	\bm f) + \bm C \bm f \, .
	\label{eq:inpainting}
\end{equation}
The discriminator $d: \R^N \times R^N \rightarrow \R$  aims to distinguish the 
distribution of the reconstruction with the known data as side information 
$\mathbb{P}(\bm u | \bm c, \bm C \bm f)$ 
from the original distribution $\mathbb{P}(\bm f | \bm c, \bm C \bm f)$, 
minimising
\begin{equation}
	\mathbb{E}_{\bm f \sim \mathbb{P}_t, \bm c \sim \mathbb{P}_{c}} 
		\left(\mathbb{E}_{r\sim\mathbb{P}_s} d(\bm u(\bm r, \bm c, \bm C \bm 
	f),  \bm c) - d(\bm f, \bm c) \right) \, .
	\label{eq:discloss}
\end{equation}
Here, $\bm f$ is a sample from the natural image distribution $\mathbb{P}_t$, 
$\bm c$ a random mask from the distribution $\mathbb{P}_c$, and $\bm r$ a 
uniformly random seed. 
$\mathbb{E}$ denotes the expected value which is estimated in practice via the 
batch mean. To approximate the Lipschitz property required by the Wasserstein 
distance, the discriminator weights are normalised to 1 in the 2-norm (see 
\cite{VHF21}).
The generator has a combined loss which weights the discriminator opinion with 
parameter $\alpha$ against a mean absolute error (MAE) in terms of the 1-norm 
$\|\cdot\|_1$ that attaches the result to the concrete original image $\bm f$:
\begin{equation}
	\begin{split}
		\mathbb{E}_{\bm f \sim \mathbb{P}_t, \bm r \sim \mathbb{P}_{\bm s}, \bm 
		c \sim 
			\mathbb{P}_{\bm c}}  \big( &-\alpha 
		d(\bm u(\bm r, \bm c, \bm C \bm f), \bm 
		c)  \\ 
		&+ 
		\| \bm f - \bm u(\bm r, \bm c, \bm C \bm f) \|_1 \, \big) \, .
	\end{split}
	\label{eq:genloss}
\end{equation}

Va\v{s}ata et al.~\cite{VHF21} use a common hourglass structure for $\bm g$ 
that successively subsamples the input data, passes it through a bottleneck and 
upsamples it again to the output (see $g$ in 
Fig.~\ref{fig:model_structure}(a)). Skip connections forward data between 
corresponding scales in this hierarchical network. The building blocks of their 
architecture are visualised in Fig.~\ref{fig:model_structure}(b)-(d).

Downsampling in the hourglass is performed by \emph{CBlocks} using 3 parallel 
convolutions with filter size $5\times5$, dilation rates $0$, $2$, and $5$, and 
ELU activation. Their concatenated output is followed by a $2 \times 2$ max 
-pooling, which is the output of the block.  Upsampling in \emph{TCBlocks} 
follows the same principle with transposed convolutions and $2 \times 2$ 
upsampling instead. The parallel dilations increase the influence area of these 
blocks, which is particularly useful for sparse known data, since it increases 
the chance to include some reliable pixels in the receptive field of the 
convolution layers. To restrict the image to the original pixel value range 
$[0,1]$, the last transposed convolutional layer has a hard sigmoid 
activation.  

The discriminator follows a simpler downsampling architecture, where 
\emph{FBlocks} combine $5\times5$ convolutions (stride $2$) with Leaky ReLUs 
(see $d$ in Fig.~\ref{fig:model_structure}). For exact details we refer to 
\cite{VHF21}.

%---------------------------------------------------------------------------

\section{Learning Masks with Wasserstein GANs}
\label{sec:method}

\subsection{Learning Binary Masks}
\label{sec:binary}

%............................................................................

\begin{table}[t]
\begin{center}
\tabcolsep15pt
\begin{tabular}{c|c|c}
additive noise & stochastic rounding & hard rounding \\
$b(x)=x + \varepsilon$ & $b(x) \in \{0,1\},$  & 
$b(x)=\lfloor x + 0.5 \rfloor$ \\
& $P(b(x)\!=\!1)\!=\!x\!-\!\lfloor x \rfloor$ &
\end{tabular}
\end{center}
\caption{\label{tab:binarisation} \textbf{Binarisation with Quantisation 
Operators.} Here, $\varepsilon$ is chosen uniformly random from 
$[0,0.5]$. Among the three different options for binarisation operators 
\cite{TSCH17}, we choose hard rounding due to its simplicity and good 
performance in our practical implementation.}
\end{table}

%............................................................................

Existing networks \cite{APW21,DCPC19}  produce a non-binary confidence map 
$\bm c \in [0,1]^N$ during training. Dai et al. \cite{DCPC19} have identified 
this as a major roadblock for joint training of an inpainting and mask 
networks. The known data $\bm C \bm f$ obtained in Eq.~\eqref{eq:inpainting} 
provides information to the inpainting network that is not available during 
actual inpainting. Therefore, we need to binarise the mask already during 
training.

Unfortunately, binarisation is non-differentiable and thus prevents  
backpropagation. We solve this issue with a \emph{binarisation block} 
for the last step of our mask generator. In a first step of this block, we 
apply a transposed convolution with a single channel output, followed by a hard 
sigmoid activation. We interpret the conversion of this non-binary output $\bm 
c \in [0,1]^N$ into a binary mask $\bm b \in \{0,1\}^N$ as an extreme case of 
quantisation. Such a discretisation of the co-domain restricts the admissible 
range of values, in our case just $0$ and $1$.  

In deep compression, non-differentiability is often addressed by choosing 
representatives of the quantisation intervals according to additive random 
noise~\cite{BLS17}. Theis et al.~\cite{TSCH17} have investigated different 
quantisation strategies in neural network-based compression and the impact of 
quantisation perturbations on training (see Table~\ref{tab:binarisation}). 
Compared with additive noise and stochastic rounding they found hard rounding 
to perform the best. Therefore, we round non-binary confidence values $c$ 
according to $b(c)=\lfloor c + 0.5 \rfloor$. Following the findings of Theis et 
al., we approximate the gradient of the binarisation layer by the derivative of 
a simple linear function for backpropagation.

%............................................................................

\begin{figure}[t]
\centering
\includegraphics[width=\textwidth]{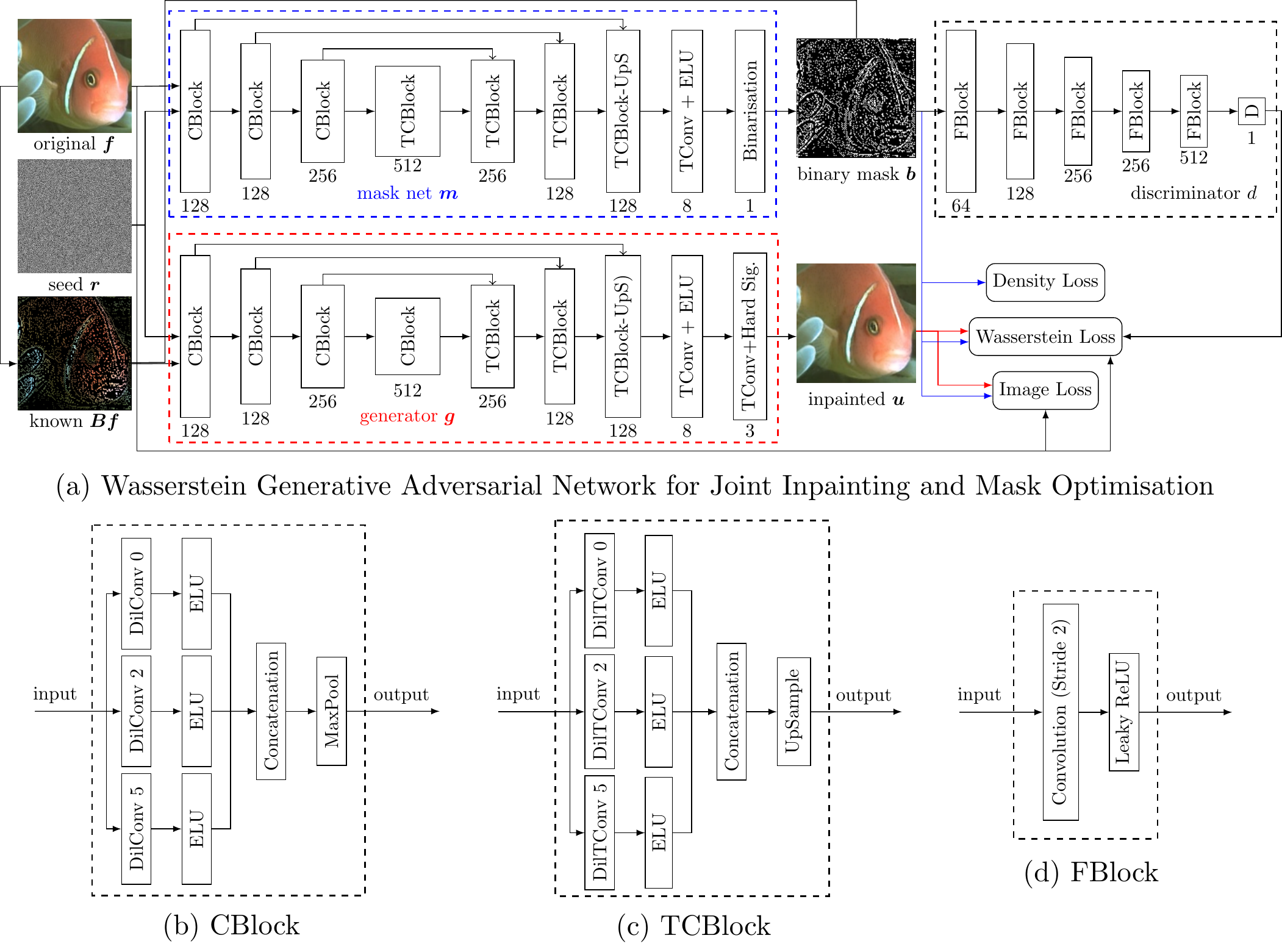}
\caption{\textbf{Overview over our model structure.} The arrows denote 
forward passes. CBlocks, TCBlocks, and FBlocks denote convolutional, 
transposed convolutional, and funnel blocks (see  
Section~\ref{sec:WassersteinGANs}). The las TCBlock omits the upsampling layer 
(indicated by -UpS). TConv denotes transposed convolutions, DConv dilated 
convolutions, and DTConv their transposed counterpart. All convolutions use 
filter size $5\times5$.
The numbers in dilated convolution layers indicate the dilation parameter.
The block height in (a) visualises resolution and the block width roughly 
indicates the number of output channels, which is precisely given by the 
numbers below each block.
\label{fig:model_structure}}
\end{figure}

%---------------------------------------------------------------------------

\subsection{Joint Learning of Inpainting Operator and Masks}
\label{sec:joint}

The overall structure of our joint mask and inpainting WGAN is displayed in 
Fig.~\ref{fig:model_structure}. Our new \emph{mask generator} $\bm m$ maps the 
original $\bm f \in \R^N$  and a uniformly random seed $\bm r \in \R^N$ to a 
binary mask $\bm b  = \bm m(\bm r, \bm f)$.
The generator $\bm g$ uses this mask and the known data $\bm B \bm f 
:= \textnormal{diag}(\bm b) \bm f$ as side information to create the inpainting 
result $\bm u$ from another random seed. The discriminator loss from  
Eq.~\ref{eq:discloss} and the generator loss from Eq.~\ref{eq:genloss} ensures 
that the inpainting respects the statistics of natural images. However, the 
mask $\bm c$ is replaced by the output $\bm b$ of the mask generator. 

Unfortunately, we have no way to obtain training data for the unknown 
distribution of the binary masks that should be approximated by the mask 
generator~$\bm m$. This distribution depends on the inpainting operator which 
is simultaneously trained, thus creating a ``chicken and egg'' problem. We 
solve this by indirectly describing the distribution: The mask generator is 
coupled to the Wasserstein loss of the discriminator and generator, since $\bm 
b$ influences the inpainting result~$\bm u$. Moreover, we define a 
\emph{density loss} that measures the deviation of percentage of known pixels 
$\|\bm b\|_1/N$ from the target density $D$ of the inpainting mask. The 
following loss is directly imposed on the mask network, weighting the density 
loss against the inpainting MAE by $\beta$:
\begin{equation}
	\begin{split}
		\mathbb{E}_{\bm f \sim \mathbb{P}_t, \bm r \sim \mathbb{P}_{\bm s}} 
		& \big( \, \big| \, \|\bm m(\bm r, \bm f)\|_1 / N - D \, \big| \\
		&+ \beta \, \| \bm f -\bm  u(\bm r, \bm m(\bm r, \bm f), \bm B \bm f) 
		\|_1 
		\big) \, .
	\end{split}
	\label{eq:maskloss}
\end{equation}

The architecture of the generator and discriminator is identical to the one 
from Section~\ref{sec:WassersteinGANs}. The mask generator mostly follows the 
inpainting generator design. We only replace the last block by the binarisation 
from Section~\ref{sec:binary}. Due to the Wasserstein loss, training is  
straightforward: In each epoch, we update the weights of all three networks 
with backpropagation. Training remains stable and requires no fine tuning of 
the balance between generator and discriminator.  

%---------------------------------------------------------------------------

\section{Experiments}
\label{sec:experiments}

\subsection{Experimental Methodology}

We compare against state-of-the-art methods for optimisation of inpainting data.
As discussed in Section~\ref{sec:related}, most data optimisation techniques 
are still model-based. Out of the various existing approaches, we choose 
probabilistic sparsification (PS) and non-local pixel exchange (NLPE) 
\cite{MHWT12}, which mark the qualitative state of the art for most inpainting 
techniques, including the widely-used homogeneous diffusion \cite{Ii62} 
inpainting. In particular, the diffusion mask network of Alt et al.~\cite{APW21}
reaches similar quality as PS on greyscale data. Results on colour images are 
not available and a corresponding extension would be non-trivial. 

Our direct competitor on the neural network side is NetM \cite{DCPC19} since it optimises known data for deep inpainting. 
Therefore, we compare against NetM in combination with all inpainting operators 
evaluated by Dai et al.-~\cite{DCPC19}. The numerous pure neural networks for 
inpainting  discussed in Section~\ref{sec:related} only perform inpainting of 
pre-defined masks and do not perform spatial optimisation. They can therefore 
not be considered for comparison.

Our neural networks were trained on the same $100,000$ image \emph{ImageNet} 
subset also used by Dai et al.~\cite{DCPC19} and the corresponding validation 
set. Depending on the evaluation set, these were centre-cropped to either $128 
\times 128$ or $64\times 64$.  We used the \emph{Adam} optimiser \cite{KB15} 
with learning rate $5\cdot10^{-5}$ and a batch size of $b=32$ for image size 
$128\times128$, and $b=128$ for image size $64 \times 64$. The model parameters 
were set to $\alpha=0.005$ and $\beta=1$. For each mask density $D$, separate 
networks were trained. We chose the best weights w.r.t. the mask validation 
loss from Eq.~\eqref{eq:maskloss}  after 1000 epochs. In most cases, this was 
already reached after roughly $100$ epochs. 

As evaluation datasets we use the ImageNet test set provided by Dai et 
al.~\cite{DCPC19} for the network comparison. For the comparison with NLPE, we 
use the Berkeley shape database \emph{BSDS500} \cite{AMFM11}, since it has 
better public availability and this also demonstrates that our networks 
transfer well to other natural image databases.

%---------------------------------------------------------------------------

\begin{table}[t]
\begin{center}
\tabcolsep3pt
\begin{tabular}{c|cc|cccc|c}
&\multicolumn{2}{|c|}{(a) random masks}&\multicolumn{4}{|c|}{(b) 
	optimised NetM masks}&\\ 
Density & NetE \cite{DCPC19} & WGAN \cite{VHF21} & NetE \cite{DCPC19} 
& CDD \cite{CS01a} & BPFA \cite{ZCPR+12} & MS \cite{ES02} &\textbf{Our MG}
\\
\hline
5\%\Tstrut & 18.44 & 18.85 & 20.35 & 20.22 & 15.82 & 20.70 & \textbf{21.66} 
\\
10\% & 19.94 & 20.78  &21.93 & 22.42 & 20.82 & 22.98 & \textbf{23.63} \\
20\% & 22.06 & 22.95 & 24.31 & 24.38 & 24.16 & 25.04 & \textbf{25.36}  \\
\hline 
\end{tabular}
\end{center}
\caption{\label{tab:netm} \textbf{PSNR Comparison Against NetM on ImageNet}  
(higher is better). Our MG approach outperforms NetM in combination with all 
four inpainting operators investigated by Dai et al.~\cite{DCPC19}. In 
particular, it outperforms the full network approach with NetE consistently by 
more than 1dB in PSNR, even though MG does not maximise PSNR.}
\end{table}

%---------------------------------------------------------------------------

\subsection{Comparison Against NetM}

We compare against the mask generator NetM \cite{DCPC19} on the test set 
curated by Dai et al. It consists of $1000$ images or resolution $64 
\times 64$. They provide peak-signal-to-noise ratio (PSNR) results of NetM in 
combination with the corresponding inpainting network NetE and multiple 
classical inpainting approaches \cite{CS01a,ES02,ZCPR+12}. Note that NetM is 
trained to minimise a 2-norm, giving it an advantage over our 
1-norm/Wasserstein trained MG in this evaluation. 
Nevertheless, in Table~\ref{tab:netm}(b), MG outperforms NetM+NetE   
substantially by up to 1.7 dB.

To verify that this advantage does not only result from using a GAN for 
inpainting, we also  compare the inpainting WGAN \cite{VHF21} to NetE on random 
masks in  Table~\ref{tab:netm}(a). On 5\% known data, the WGAN outperforms NetE 
only by 0.4 dB. Our mask GAN increases this improvement over NetM+NetE to 1.3 
dB. This indicates that our mask binarisation and joint training offers an 
advantage.

Additional inpainting operators in combination with NetM, such as the \\
{Bayesian} beta process factor analysis (BPFA) \cite{ZCPR+12}, curvature-driven 
diffusion (CCD) \cite{CS01a}, or Mumford-Shah (MS) inpainting \cite{ES02} all 
yield worse results than our mask GAN. We outperform the best NetM approach by 
up to 0.96 dB.

\subsection{Comparison Against Probabilistic Methods}

Probabilistic sparsification (PS) with a non-local pixel exchange (NLPE) as a 
postprocessing step defines a benchmark for the best results obtainable so far 
with homogeneous diffusion inpainting. Other methods 
\cite{APW21,BBBW08,BLPP16,CRP14,HSW13} yield comparable or worse quality. 

We optimise the stochastic models for mean average error (MAE) since this is 
also part of the network loss. NLPE uses 5 cycles of $|\bm c|$ iterations. At 
first glance, in terms of MAE, our mask GAN (MG) is situated 
between PS and PS+NLPE in Table~\ref{tab:nlpe}(a). It performs better on 
sparser masks which are more relevant for e.g.~compression applications and 
comes very close to the NLPE error for 5\%. We also evaluate w.r.t. the popular 
structural-similarity index (SSIM) \cite{WBSS04} in Table~\ref{tab:nlpe}(b). 
It yields a slightly different ranking, with our mask GAN also outperforming 
PS+NLPE on the $5\%$ density and very similar values for all methods on $10\%$.

%---------------------------------------------------------------------------

\begin{table}[t]
\tabcolsep4pt
\begin{center}
\begin{tabular}{c|ccc|ccc}
	& \multicolumn{3}{|c|}{(a) MAE (lower is better)} &  
	\multicolumn{3}{|c}{(b) SSIM 
		(higher is better)}\\
	Density & PS & PS+NLPE & \textbf{Our MG} & PS & PS+NLPE & 
	\textbf{Our 
		MG} \\
	\hline
	5\% & 13.98 & 10.98 & 11.19 & 0.69 & 0.70 & 0.71 \\ 
	10\% & 9.08 & 7.51 &  8.87 & 0.80 & 0.81 & 0.79  \\
	20\% & 5.19 &  4.31 & 6.70 & 0.90 & 0.91 & 0.85  \\
	\hline 
\end{tabular}
\end{center}
\caption{\label{tab:nlpe} \textbf{Quantitative Comparison to Stochastic Methods 
on BSDS500.}\\ (a) W.r.t. MAE, our mask GAN (MG) outperforms probabilistic 
sparsification (PS) consistently and is competitive with non-local pixel 
exchange (NLPE) on low densities. (b) In terms of SSIM, at low densities the 
mask GAN slightly outperforms both competitors while remaining competitive for 
higher densities at significantly reduced computational load (see Table 
\ref{tab:runtime}).}
\end{table}

%---------------------------------------------------------------------------

\begin{figure}[p]
\centering
\tabcolsep1pt
\def\picsize{0.240}
\begin{tabular}{cccc}
Original &	 PS \cite{MHWT12} & PS+NLPE \cite{MHWT12} & \textbf{Our MG} 
\\
\includegraphics[width=\picsize\textwidth]{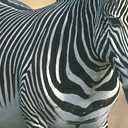}\hspace{0.4mm}
&
\includegraphics[width=\picsize\textwidth]{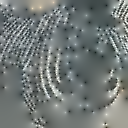}\hspace{0.4mm}
&
\includegraphics[width=\picsize\textwidth]{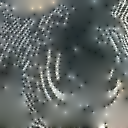}\hspace{0.4mm}
&
\includegraphics[width=\picsize\textwidth]{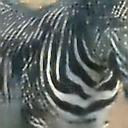}  \\
(a) image 130066
& MAE 41.53 & MAE 36.25 & MAE 27.32 \\
& SSIM 0.35 & SSIM 0.34 & SSIM 0.59 \\[1mm]
&\includegraphics[width=\picsize\textwidth]{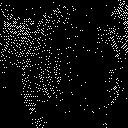}
& 		
\includegraphics[width=\picsize\textwidth]{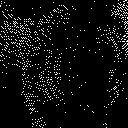}
&
\includegraphics[width=\picsize\textwidth]{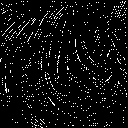}  \\
& 902 pixels  & 902 pixels  & 902 
pixels \\[1mm]
\includegraphics[width=\picsize\textwidth]{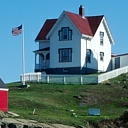}\hspace{0.4mm}
&		
\includegraphics[width=\picsize\textwidth]{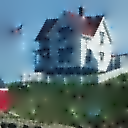}\hspace{0.4mm}
&	
\includegraphics[width=\picsize\textwidth]{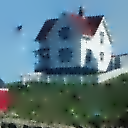}\hspace{0.4mm}
&	
\includegraphics[width=\picsize\textwidth]{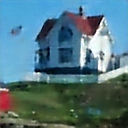}  \\
(b) image 228076 &  MAE 18.39 &  MAE 14.62 & MAE 12.77 \\
&  SSIM 0.65 &  SSIM 0.67 & SSIM 0.69 \\
&	
\includegraphics[width=\picsize\textwidth]{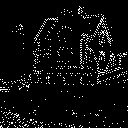}
&
\includegraphics[width=\picsize\textwidth]{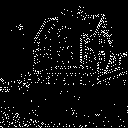}
&
\includegraphics[width=\picsize\textwidth]{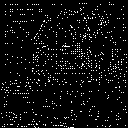}
\\
& 846 pixels  & 846 pixels & 846 pixels  \\		
\end{tabular}
\caption{\label{fig:visual} \textbf{Visual Comparison to Probabilistic 
Methods at $\approx 5\%$ Density on BSDS500.} Known data points are marked in 
white. At this extreme density that might be used for e.g. thumbnail  
compression or destructive image acquisition, our mask WGAN reconstructs  
structures from less clustered data. It does not suffer from visual artefacts 
such as singularities or extreme colour bleeding as PS and NLPE with 
homogeneous diffusion.
}
\end{figure}

Since these error measures yield less clear quantitative results than our first 
set of experiments, we also provide multiple visual examples. 
Fig.~\ref{fig:visual} demonstrates that our MG excels for low densities which 
are useful for applications such as thumbnail compression or destructive image 
acquisition. 
This holds especially for complex images, for instance the high contrast 
texture of the zebra, or the house with many small-scale details like the lawn 
and fence. PS and NLPE need to cluster known data left and right to edges 
according to the optimality theory of Belhachmi et al.~\cite{BBBW08}. In 
regions where this is not possible, they suffer from detail loss and colour 
bleeding. Sometimes, minimising the MAE with NLPE even leads to a slight 
deterioration w.r.t. SSIM, since some edges are reinforced while others vanish 
(see Fig.~\ref{fig:visual}(a) and Fig.~\ref{fig:visual}(b)). In contrast, our 
approach can reconstruct structural image features from less known data. 
Therefore, it can distribute the mask pixels much more evenly.

%---------------------------------------------------------------------------

%---------------------------------------------------------------------------

\begin{figure}[p]
\centering
\tabcolsep1pt
\def\picsize{0.240}
\begin{tabular}{cccc}
Original &	 PS \cite{MHWT12} & PS+NLPE \cite{MHWT12} & \textbf{Our MG} 
\\
\includegraphics[width=\picsize\textwidth]{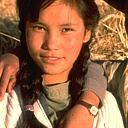}\hspace{0.4mm}
&
\includegraphics[width=\picsize\textwidth]{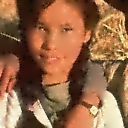}\hspace{0.4mm}
&
\includegraphics[width=\picsize\textwidth]{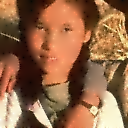}\hspace{0.4mm}
&
\includegraphics[width=\picsize\textwidth]{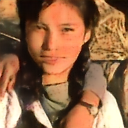}  \\
(a) image 376001 &  MAE 12.88 &  MAE 11.09 & MAE 11.03 \\
& SSIM 0.79 & SSIM 0.78 & SSIM 0.82 \\[1mm]
&	
\includegraphics[width=\picsize\textwidth]{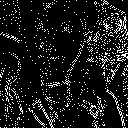}
&
\includegraphics[width=\picsize\textwidth]{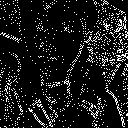}
&
\includegraphics[width=\picsize\textwidth]{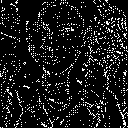}
\\
& 1667 pixels  & 1667 pixels & 1668 pixels \\[1mm]
\includegraphics[width=\picsize\textwidth]{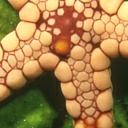}\hspace{0.4mm}
&
\includegraphics[width=\picsize\textwidth]{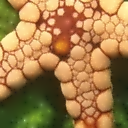}\hspace{0.4mm}
&
\includegraphics[width=\picsize\textwidth]{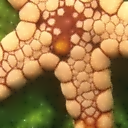}\hspace{0.4mm}
&
\includegraphics[width=\picsize\textwidth]{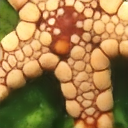} \\
(b) image 12003 & MAE 5.73 & MAE 5.22 & MAE 6.16 \\
& SSIM 0.93 & SSIM 0.94 & SSIM 0.91 \\[1mm]
&\includegraphics[width=\picsize\textwidth]{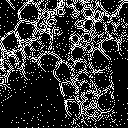}
& 		
\includegraphics[width=\picsize\textwidth]{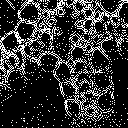}
&
\includegraphics[width=\picsize\textwidth]{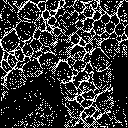}  \\
& 3234 pixels  & 3234 pixels  & 3234 pixels \\
\end{tabular}
\caption{\label{fig:visual2} \textbf{Visual Comparison to Probabilistic 
Methods at 10\% and 20\% on BSDS500.} Known data points are marked in white. 
With increasing density, the quality of all three methods approaches each 
other. In (a), our MG still benefits from better known data distribution, 
yielding a more detailed representation of the face. In~(b), it produces a 
similar visual quality as PS and NLPE at a higher error. It does not suffer 
from singularities, but this is less notable since the quality is overall high.
}
\end{figure}

%---------------------------------------------------------------------------

These advantages are less pronounced for higher densities or simpler images. 
With higher amounts of known data, homogeneous diffusion can reconstruct more 
edges and all approaches become more similar w.r.t. quality. While the MAE and 
SSIM scores for high densities are slightly worse on average for our mask GAN 
in Table~\ref{tab:nlpe}, its visual quality is competitive. The Wasserstein 
loss does not lead to the smallest possible quantitative errors, but yields 
natural looking inpainting results in Fig.~\ref{fig:visual2}. In particular, 
our mask approach does not suffer from the visually unpleasant singularities of 
homogeneous diffusion inpainting that are clearly visible in 
Fig.~\ref{fig:visual2}(a). Here we can also observe that its more even error 
distribution often leads to better representations of visually important image 
content such as faces.

In addition to its competitive visual quality, our Mask GAN is   
substantially faster than a PS/NLPE conjugate-gradient implementation with 
relative residual $10^{-6}$.  On a single CPU core, the speed-up reaches up to 
a factor $\approx 63$ w.r.t. PS and a factor $\approx 419$ w.r.t. NLPE. With 
GPU support, our GAN can be  up to $\approx 12,560$ times faster. 
Faster model-based alternatives to PS exist \cite{BBBW08,CW21}, but so far, 
they typically require postprocessing to reach the quality of PS+NLPE. Since 
they use homogeneous diffusion, they also suffer from visual artefacts like 
singularities. Overall, our network provides a fast solution
for sparse data optimisation with a high visual quality.

%---------------------------------------------------------------------------

\begin{table}[t]
\begin{center}
\tabcolsep5pt
\begin{tabular}{ccccc}
	Density & PS & PS+NLPE &  \textbf{Our MG (CPU)} & \textbf{Our MG 
	(GPU)} 
	\\
	\hline
	5\% & 58.84 & 251.28 & 0.93 & 0.031 \\
	10\% & 33.89 & 340.61 & 0.93 & 0.031  \\
	20\% & 18.86 & 389.34 & 0.93 & 0.031    \\
	\hline 
\end{tabular}
\end{center}
\caption{\label{tab:runtime} \textbf{Runtime Comparison} on $128 \times 
128$ images with an Intel Core i56660K@3.50GhZ and a NVIDIA Geforce GTX 1070. 
Our WGAN is faster than classical methods by several orders of magnitude 
(factor $>400$ on CPU or $>12,000$ on GPU). Its runtime is independent of the 
mask density.}
\end{table}

%---------------------------------------------------------------------------

\section{Conclusion and Future Work}
\label{sec:conclusion}

We have presented the first adversarial network for joint learning of a 
generative inpainting operator and a binary mask generator. Based on a 
mathematically well-founded Wasserstein framework, our inpainting GAN 
approximates the statistics of natural images, yielding visual improvements 
over model-based stochastic approaches with homogeneous diffusion. 
Simultaneously, our approach is faster by several orders of magnitude. It also 
qualitatively outperforms competing neural networks for spatial optimisation in 
combination with many inpainting operators. 

Currently, we are working on further refinements of both the general framework 
and the concrete network architecture. Moreover, we plan to evaluate the impact 
of individual components with an extended evaluation and ablation study in the 
future. Our model is a step towards fast, visually accurate, and 
mathematically justified spatial optimisation with deep learning. We hope 
that it contributes to practical applications such as image compression or 
adaptive sampling in the future.

\subsubsection{Acknowledgements} 
This work has received funding from the European Research Council 
(ERC) under the European Union's Horizon 2020 research and 
innovation programme (grant agreement no. 741215, ERC Advanced 
Grant IN\-CO\-VID).
We thank Dai et al. \cite{DCPC19} for providing their reference dataset.

\bibliographystyle{splncs04}
%\bibliography{myrefs}

%

\end{document}